\documentclass[useAMS,usegraphicx]{mn2e} 

\newcommand{\Teff}{\mbox{$T_{\mbox{\scriptsize eff}}\,$}}
\newcommand{\Rsolar}{\mbox{R$_{\odot}$}}
\newcommand{\Msolar}{\mbox{M$_{\odot}$}}

\newcommand{\kms}{\mbox{${\rm km\,s}^{-1}$}}

\newcommand{\logg}{\mbox{$\log g$}}
\newcommand{\rxj}{\mbox{\rm RX\,J2130.6+4710}}

\title[RR Cae]
{The mass and radius of the M-dwarf in the short period eclipsing binary
RR Caeli }
\author[P.~F.~L. Maxted et~al.]
{P.~F.~L. Maxted$^{1,2}$,  D. O'Donoghue$^{3}$, L. Morales-Rueda$^{4,2}$, 
R. Napiwotzki$^{5}$,
\newauthor B. Smalley$^{1}$ \\
 $^1$Astrophysics Group, Keele University, Staffordshire, ST5~5BG,
 UK. \\
 $^2$University of Southampton, Department of Physics \& Astronomy,
 Highfield, Southampton, S017 1BJ, UK. \\
 $^3$South African Astronomcal Observatory, PO Box 9, 
Observatory 7935, Cape Town, South Africa. \\
 $^4$Department of Astrophysics,  University of Nijmegen,
  P.O.~Box~9010, 6500 GL Nijmegen, The Netherlands. \\
$^{5}$Centre for Astrophysics Research, STRI, University of Hertfordshire,
Hatfield, AL10 9AB, UK}

\date{Accepted --- Received ---}

\pagerange{\pageref{firstpage}--\pageref{lastpage}}

\pubyear{2006}

\begin{document}

\maketitle

\label{firstpage}
\begin{abstract} 
 We present new photometry and spectroscopy of the eclipsing white dwarf --
M-dwarf binary star RR~Cae. We use timings of the primary eclipse from
white-light photo-electric photometry to derive a new ephemeris for the
eclipses. We find no evidence for any period change greater than $|\dot{P}|/P
\approx 5\times 10^{-12}$ over a timescale of 10 years. We have measured the
effective temperature of the white dwarf, T$_{\rm WD}$, from an analysis of
two high resolution spectra of RR~Cae and find  T$_{\rm WD} = 7540\,{\rm K}\pm
175\,{\rm K}$. We estimate a spectral type of M4 for the companion from the
same spectra. We have measured the radial velocity of the white dwarf from the
Balmer absorption lines and find that the semi-amplitude of the spectroscopic
orbit is $K_{\rm WD} = 79.3 \pm 3.0$\,\kms. We have combined our radial
velocity measurements of the M-dwarf with published radial velocities to
determine a new spectroscopic orbit for the M-dwarf with a semi-amplitude of
$K_{\rm M} = 190.2 \pm 3.5$\,\kms. We have combined this information with an
analysis of the primary eclipse to derive relations between the inclination of
the binary and the radii of the two stars. We use cooling models for helium
white dwarfs with a wide range of hydrogen layer masses to determine the
likely range of the white dwarf radius and, thus, the inclination of the
binary and the mass and radius of the M-dwarf. The mass of the M-dwarf is
(0.182\,--\,0.183)$\pm$0.013\,\Msolar\ and the radius is
(0.203\,--\,0.215)$\pm$0.013\,\Rsolar,  where the ranges quoted for these
values reflect the range of white dwarf models used. In contrast to previous
studies, which lacked a measurement of $K_{\rm WD}$,  we find that the mass
and radius of the M-dwarf are normal for an M4 dwarf. The mass of the white
dwarf is 0.440 $\pm$0.022\,\Msolar. With these revised masses and radii we
find that RR~Cae will become a cataclysmic variable star when the orbital
period is reduced from its current value of $7.3$\,hours to 121\,minutes by
magnetic braking in 9\,--\,20\,Gyr. We note that there is night-to-night
variability of a few seconds in the timing of primary eclipse  caused by
changes to the shape of the primary eclipse. We speculate as to the possible
causes of this phenomenon.

\end{abstract} 


\section{Introduction}

 RR~Cae is a cool white dwarf star which is eclipsed by  an M-dwarf companion
every 7.3\,h. The eclipses were first reported by Krzeminski (1984) in a
short report which also mentions that the ``radial-velocity amplitude
[\,of the dMe star\,] throughout the orbit is 370 km/s''. An  R-band
lightcurve has been presented by Bruch \& Diaz (1998) together with a single
spectrum covering the wavelength region 5806--7116\AA\ from which they derive
a spectral type for the M-dwarf companion of M5V or M6V. A more complete
history of the study of this star can be found therein. Bruch (1999) presented
further spectroscopy around the H$_\alpha$ line from which he derives a
spectroscopic orbit for the M-dwarf. This information was combined with an
analysis of the R-band lightcurve to estimate the masses and radii of the
stars. Both Bruch \& Diaz and Bruch estimated a mass of about 0.09\Msolar\ for
the M-dwarf, which is much lower than expected given its spectral type.
Zuckerman et~al. (2003) present a single spectrum of RR~Cae taken with the
{\sc hires} echelle spectrograph on the Keck telescope. This spectrum shows
several narrow absorption features at wavelengths around 390nm due to neutral
metals, e.g., Al\,{\sc i}, Fe\,{\sc i} and Mg\,{\sc i}, which are thought to
have been accreted onto the surface of the white dwarf from the M-dwarf.

 RR~Cae is a double-lined eclipsing binary so it
is a candidate for an accurate determination of the masses and radii of both
stars. Only a few double-lined eclipsing binaries containing white
dwarf stars have been identified. V471~Tau in the Hyades is perhaps the best
known and longest studied such binary. The K-dwarf companion dominates the
optical light of the binary so O'Brien, Bond \& Sion (2001) used Hubble Space
Telescope (HST) spectra of the Ly\,$\alpha$ line to measure the temperature,
surface gravity and spectroscopic orbit of the DAO white dwarf. They combined
these data with observations from ground-based observatories to measure the
masses and radii of the DAO and K-dwarf stars to an accuracy of about
7\,percent. They find that the DAO star has the radius expected for its mass
but that the K-dwarf is 18\,percent larger than normal Hyades dwarfs of the
same mass. They attribute this expansion to the large degree of coverage of
the stellar surface by starspots. They find that the white dwarf in V471~Tau
is the most massive one known in the Hyades but also the hottest and youngest,
in direct conflict with expectation. They suggest that  V471~Tau may be the
result of the evolution of a triple star system.

O'Donoghue et~al. (2003) report the discovery of eclipses in the white-dwarf
-- M-dwarf binary EC\,13471$-$1258. They also used HST observations combined
with observations from ground-based observatories to measure the masses and
radii of the stars. They were able to achieve an accuracy of 5--10\, percent
in these measurements despite the complications that arise from the extreme 
flaring behaviour of the M-dwarf. They find that the radius of the white dwarf
is that expected for a carbon-oxygen white dwarf and that the M-dwarf also
has the same radius as normal dwarf stars of the same mass.  O'Donoghue et~al.
argue that the M-dwarf just fills its Roche lobe and that this, combined with
the rapid rotation of the white dwarf, strongly suggests that the system has
undergone mass transfer in the past, and imply that it is a hibernating
cataclysmic variable (CV).

Maxted et~al., (2004) report the discovery of eclipses in the white-dwarf --
M-dwarf binary RX\,J2130.6+4710. They present lightcurves and spectroscopy
from which they derive masses to an accuracy of about 5\,percent and radii to
an accuracy of about 10\,percent for both stars. There is no secondary eclipse
visible in the lightcurves of this binary so Maxted et~al. used the amplitude
of the ellipsoidal effect as an additional constraint in their analysis to
resolve the degeneracy between inclination and the ratio of radii of the
stars. They find that the radius of the white dwarf is that expected for a
carbon-oxygen white dwarf and that the M-dwarf also has the same radius as
normal dwarf stars of the same mass.

There are several other white-dwarf -- M-dwarf binaries known, some of which
are eclipsing binaries (Hillwig, Honeycutt \& Robertson 2000). These tend to be
systems in which the light from the secondary star is dominated by
reprocessed radiation from the hot white dwarf. This light is asymmetrically
distributed on the surface of the M-dwarf and often produces a spectrum
dominated by emission lines, so some model of the ``reflection effect'' is
required to infer dynamical masses from the the spectroscopic orbit. It is
usually the case for these binaries that there is insufficient information to
determine the masses  of the stars without assuming that one of the
stars has a normal radius for its mass. Alternatively or additionaly, it
may be  necessary to use a surface gravity estimate for the white dwarf
derived by fitting synthetic spectra to the observed spectrum to determine the
geometry of the binary.

Schreiber \& G\"{a}nsicke (2003) have made a compilation of measured and
estimated masses, radii and effective temperatures for white-dwarf -- M-dwarf
binaries and other post common-envelope binaries. They considered the past and
future evolution of these binaries based on these parameters. They find that
the sample of known post common-envelope binaries is very incomplete. In
particular,  the majority of post common-envelope binaries that will evolve
into CVs within a Hubble time (pre-CVs) are missing from existing surveys
because the white dwarf is too cool to stand out from it's companion star.
Much of this analysis is based on indirect estimates of the masses and radii
mentioned above. One reason to study binaries like RR~Cae, RX\,J2130.6+4710,
EC\,13471$-$1258 and V471~Tau is to test the accuracy of the methods used to
make these indirect estimates.

 In this paper we present new spectroscopy and photometry of RR~Cae. We derive
the spectroscopic orbit of the white dwarf and determine the mass ratio of the
binary directly. We present photometry in white-light of the eclipse from which
we derive an accurate ephemeris. We combine an analysis of our white-light
data and spectroscopic orbits with cooling models for the white dwarf to
measure the mass and the radius of the M-dwarf in RR~Cae. In contrast to
previous studies, we find that the radius and mass of the M-dwarf are normal
for a star of this spectral type.

\section{Observations \& Reductions}

\subsection{Photometry}
 
We observed RR~Cae with a photomultiplier photometer on the 0.75-m and 1.0-m
telescopes at the South African Astronomical Observatory (SAAO) near
Sutherland, South Africa. These data were obtained unfiltered (i.e., in
``white-light'') with 1-s continuous integrations. The blue sensitive
photocathode of the detector thus provides a very broad response equivalent to
the combination of Johnson U+B+V passbands. The data were reduced by
subtracting sky and correcting for extinction using an appropriate mean
extinction coefficient.

 We also observed RR~Cae using the SAAO 0.75-m telescope using the UCT CCD
photometer. The UCT CCD is a Wright Instruments Peltier-cooled camera with a
576$\times$420 thinned, back-illuminated EEV charge-coupled device (CCD). The
plate scale of the resulting images is 0.37\,arcsec per pixel. Observations
were obtained on 11 nights in the interval 1999 December 7\,--\,20. We first
subtracted the bias value calculated from overscan regions from all the
images. We used the median value of several images of the twilight sky devoid
of bright stars to calculate a flat-field calibration image which was then
normalized and applied to all the images of RR~Cae. We used optimal photometry
(Naylor 1998) to determine instrumental magnitudes of the stars in each frame.
There are three comparison stars in the images we obtained, althought they are
all 3--4 magnitudes fainter than RR~Cae in the I-band. In total, we obtained
over 3000 images of RR~Cae. There are clear differences from night-to-night in
the differential magnitudes we have calculated. This is due to a
combination of several effects, including intrinsic variability of
RR~Cae, differential extinction between RR~Cae and the
comparison stars and inaccuracies in the flat-fielding. This latter effect is
particularly acute for our UCT CCD images because the twilight sky is a
different colour to the night sky and the stars themselves. This results in
systematic errors that vary with the position of the stars in the CCD images
and the airmass. For these reasons we have only used these data to measure the
depth and times of primary eclipse.

\subsection{Spectroscopy}

Observations of RR~Cae were obtained with the grating spectrograph on the SAAO
1.9-m Radcliffe Telescope. The detector was a SITe charge-coupled device (CCD)
with 1798 pixels 15$\mu$\,m wide in the direction of dispersion.  We used a
1200 line/mm grating and a 1.5\,arcsec slit to obtain 66 spectra with a
resolution of about 1\AA\ covering the wavelength region 4050\,--\,4915\AA\ at
a dispersion of 0.5\AA/pixel. Observations were obtained on the nights 1999
December 15\,--\,19. We used the same instrument to obtain 15 spectra with a
resolution of about 0.85\AA\ covering the wavelength region 8365\,--\,8975\AA\
at a dispersion of 0.35\AA/pixel on the night 1999 December 20. The
signal-to-noise ratio in our spectra is typically 5\,--\,10.

 Extraction of the narrow-slit spectra from the images was performed
automatically using optimal extraction to maximize the signal-to-noise of the
resulting spectra (Marsh 1989). The arcs associated with each stellar spectrum
were extracted using the same weighting determined for the stellar image to
avoid possible systematic errors due to the tilt of the spectra on the
detector.  The wavelength scale was determined from a fourth-order polynomial
fit to measured arc line positions. The standard deviation of the fit to the 
arc lines was typically 1/10 of a pixel. The wavelength scale for an individual
spectrum was determined by interpolation to the time of mid-exposure from the
fits to arcs taken before and after the spectrum to account for the small
amount of drift in the wavelength scale ($<0.2$\AA) due to flexure of the
instrument between arc spectra. Statistical errors on every data point
calculated from photon statistics are rigorously propagated through every
stage of the data reduction.

\section{Analysis}
\subsection{Ephemeris \label{EphemSect}}
\begin{figure}
\includegraphics[width=0.45\textwidth]{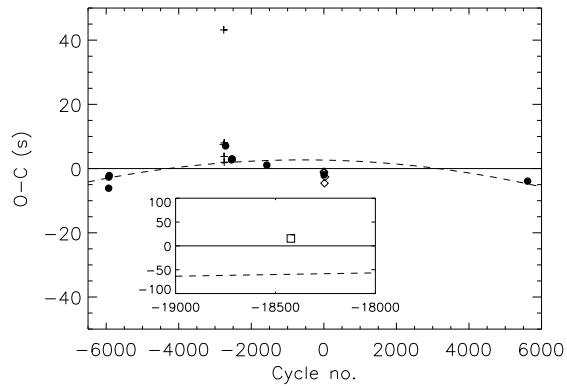}
\caption{\label{EphemFig}Residuals from a linear ephemeris for our measured
times of mid-eclipse. Symbols are as follows: dots -- white-light data from
this paper; crosses -- Bruch \& Diaz (1998); open diamonds -- I-band
data from this paper; open square (inset, note change of scale) -- Krzeminski
(1984). Note that only the white-light data are used to determine the
ephemeris. The difference between  the times of minimum predicted by our
quadratic and a linear ephemerides is shown using a dashed line. }
\end{figure}

\begin{table}
\caption{Times of mid-eclipse for RR~Cae. The source of the data are indicated
as follows: R -- Bruch \& Diaz (1998); W -- white-light data from this paper,
I -- I-band data from this paper; K -- Krzeminski (1984). The cycle number and
residual from our linear ephemeris, O$-$C, are also given. Note that only  the
white-light data are used to determine the ephemeris. \label{TminTable}}
\begin{center}

\begin{tabular}{@{}lrrc}
\multicolumn{1}{@{}l}{BJDD(Mid-eclipse)} &
\multicolumn{1}{l}{Cycle} &
O$-$C (s)& Source \\
\noalign{\smallskip}
\hline
 2445927.91665              &$   -18423 $ &$   15.6$ & K \\
 2449721.478524 $\pm$ 0.000003 &$    -5932 $ &$   -6.1$ & W \\
 2449722.389675 $\pm$ 0.000003 &$    -5929 $ &$   -2.7$ & W \\
 2449726.337828 $\pm$ 0.000004 &$    -5916 $ &$   -2.2$ & W \\
 2450681.78958               &$    -2770 $ &$    7.5$ & R \\
 2450684.82703               &$    -2760 $ &$   43.2$ & R \\
 2450687.86361               &$    -2750 $ &$    3.8$ & R \\
 2450688.77470               &$    -2747 $ &$    2.0$ & R \\
 2450688.77477               &$    -2747 $ &$    8.0$ & R \\
 2450700.619201 $\pm$ 0.000004 &$    -2708 $ &$    7.1$ & W \\
 2450750.426547 $\pm$ 0.000002 &$    -2544 $ &$    2.7$ & W \\
 2450753.463587 $\pm$ 0.000002 &$    -2534 $ &$    3.0$ & W \\
 2450756.500622 $\pm$ 0.000002 &$    -2524 $ &$    2.9$ & W \\
 2451045.626463 $\pm$ 0.000002 &$    -1572 $ &$    1.1$ & W \\
 2451523.35226 $\pm$ 0.00003 &$        1 $ &$   -0.9$ & I \\
 2451524.56706 $\pm$ 0.00005 &$        5 $ &$   -1.9$ & W \\
 2451524.56707 $\pm$ 0.00003 &$        5 $ &$   -1.3$ & I \\
 2451528.51518 $\pm$ 0.00004 &$       18 $ &$   -4.5$ & I \\
 2451532.46335 $\pm$ 0.00003 &$       31 $ &$   -2.6$ & I \\
 2453228.648145 $\pm$ 0.000002 &$     5616 $ &$   -3.9$ & W \\
\noalign{\smallskip}
\hline
\end{tabular}

\end{center}    
\end{table}     

 We have used our white-light photometry to establish an ephemeris for the
times of primary minimum. We measured the time of mid-eclipse from the
lightcurve around each eclipse observed  using a least-squares fit of a very
simple lightcurve model. The model is based on the eclipse of one circular
disc with uniform surface brightness by another circular disc. 
  The measured times of minimum and their standard errors are given in
Table~\ref{TminTable}.  The precision of each measurement was calculated from
the covariance matrix of the least-squares fit. Times are reported as
Barycentric Julian date on the Barycentric Dynamical Time (TDB) timescale,
which we refer to as Barycentric Julian Dynamical date (BJDD). Corrections
from UTC to BJDD were calculated using routines from {\sc slalib} version
2.4-5 (Wallace 2000). The corrections vary from (UTC$-$TDB$)=61.7$\,s to
64.4\,s for these data.

 We found that the $\chi^2$ statistic for a least-squares fit of a linear
ephemeris was much larger than expected for a good fit ($\chi^2=2357$ cf. an
expected value of 8). This problem is reduced but not removed by using a
quadratic ephemeris ($\chi^2=125$ c.f. an expected value of 7). The poor fit
is caused by variations of a few seconds in the observed time of minimum
compared to the predicted time. We defer a discussion of the source of this
variabilty  to  Section~\ref{Discussion}, but note here that this is a
night-to-night variation, not a long term trend. To allow for this additional
source of noise in the eclipse timings we added an `external error',
$\sigma_{\rm ext}$, in quadrature to the standard error values given in
Table~\ref{TminTable}. The value of $\sigma_{\rm ext}=4.2$\,s was chosen to
give a reduced $\chi^2$ value close to 1 for the least-squares fit to the data
used to establish the following linear ephemeris:
\begin{eqnarray*} {\rm BJDD(Mid-eclipse)} = (2\,451\,523.048\,567 \pm
0.000\,019) \\ +(0.303\,703\,6366 \pm 0.000\,000\,0047   ) {\rm E}.
\end{eqnarray*}
 All phases quoted in this paper have been calculated with this ephemeris. The
residuals from this ephemeris are shown in Fig.~\ref{EphemFig} and tabulated
in Table~\ref{TminTable}.

We also used a least-squares fit to the same data to determine  a quadratic
ephemeris. We used a value of $\sigma_{\rm ext}=3.8$\,s to give a reduced
$\chi^2$ value close to 1.  
\begin{eqnarray*}
{\rm BJDD(Mid-eclipse)} = (2\,451\,523.048\,560 \pm 0.000\,018) \\
                  +(0.303\,703\,6340 \pm 0.000\,000\,0035   ) {\rm E}  \\
                  - (2.27 \pm  0.79)\times 10^{-12} {\rm E}^2
\end{eqnarray*}
We are not confident that this is a significant detection of a period change
in RR~Cae because the level of significance depends on having reliable
standard errors for the observed times of minimum, which is clearly not the
case here, and the value of the $E^2$ is strongly dependent on the single
observation on JD~2453228. However, we can confidently state that the period
of RR~Cae has not changed by more than $|\dot{P}|/P \approx 5\times10^{-12}$
over a timescale of 10 years.

Other observed times of minimum and their residuals from the linear ephemeris
presented above are given in Table~\ref{TminTable}. We used our simple
lightcurve model to measure the time of mid-eclipse from our I-band
photometry. 
We measured the full width at half minimum depth of the
eclipse from our mean white-light lightcurve and found a value of $0.027254
\pm 0.000001$ in phase units. We  used this value to convert the times of
mid-ingress and mid-egress given by Bruch \& Diaz (1998) to times of
mid-eclipse.  We have
experimented with fitting all these data to determine an ephemeris but find
that this makes a negligible change to the results presented above.
One early time of minimum is given by Krzeminski (1984).
 The uncertainty in this value is not available so it is not
clear whether it provides a useful constraint on $|\dot{P}|/P$.

\subsection{The spectroscopic orbit of the white dwarf \label{SBWD}}

\begin{table}
\begin{center}
\caption{Circular orbit fits. All velocities are in units of km/s. N is the
number of observations included in the least-squares fit. The two values of N
and $\sigma$ given  in the
column headed ``M-dwarf'' refer to the radial velocity measurements of the
absorption line spectrum presented here and from Bruch (1999), respectively.
The least-squares fit accounts for the effect of the exposure time on the
value of $K$ derived.
\label{RVFitTable}}
\begin{tabular}{@{}lrrr}
&\multicolumn{1}{l}{White dwarf} &
\multicolumn{1}{l}{M-dwarf} &
\multicolumn{1}{l}{Emission lines} \\
\hline
\noalign{\smallskip}
$\gamma$  & 81.5 $\pm $ 2.1\parbox{0mm}{$^a$} & 85.8 $\pm $ 3.6
& 80.1 $\pm $ 1.0   \\ 
K & $79.3 \pm 3.0 $ & 190.2 $\pm$ 3.5  & 196.3 $\pm $ 1.4   \\
$\sigma$ & 9.0 & 7.3, 11.9  &  4.6 \\
N & 20 & 14,21 & 20 \\
\hline
\noalign{\smallskip}
\end{tabular}
\end{center}
{$^a$}See text for a discussion of the reliability of this estimate.
\end{table}

\begin{figure*}
\includegraphics[width=0.3\textwidth]{TrailHbetaFigSpec.ps}
\includegraphics[width=0.3\textwidth]{TrailHbetaFigFit.ps}
\includegraphics[width=0.3\textwidth]{TrailHbetaFigRes.ps}
\caption{Trailed grey-scale spectra of the H$_\beta$ line in RR~Cae.
Left-to-right: observed, phase binned spectra; least-squares fit; residuals
from the fit. For the spectra, white is an intensity value of 1.25, black is
0.25. For the residuals the intensity scale is $-0.2$ to $+0.2$ from white to
black. \label{TrailHbetaFig} }
\end{figure*}

\begin{figure}
\includegraphics[angle=270,width=0.45\textwidth]{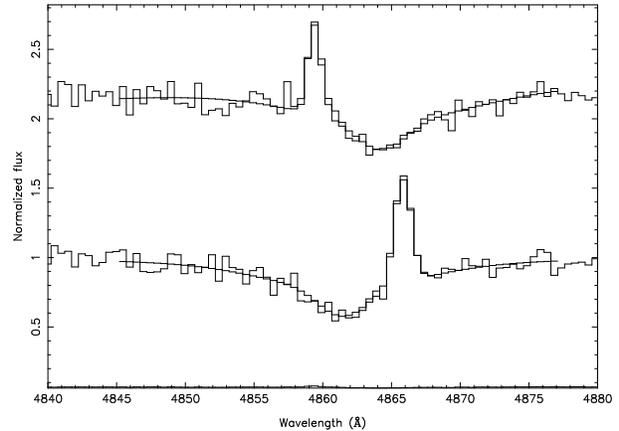}
\caption{Examples of multiple Gaussian  least-squares fits to the H$_\beta$
line. Two spectra of RR~Cae are shown (thin line)
together with the model spectra (thick line). The upper spectrum and fit have
been offset vertically by 1.2 units for clarity.
\label{SpecFitFig}} \end{figure}

\begin{figure}
\includegraphics[width=0.45\textwidth]{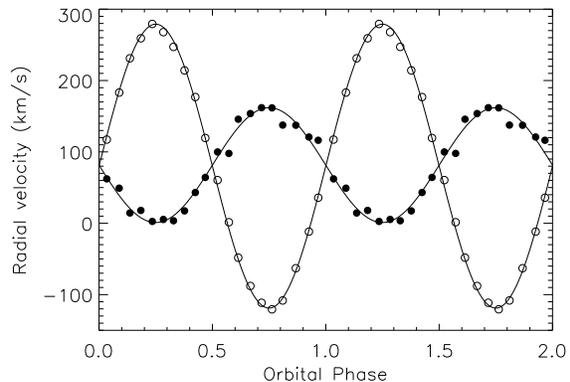}
\caption{Measured radial velocities for the white dwarf (filled circles) and
Balmer emission lines (open circles). \label{BlueRVFitFig}}
\end{figure}


\begin{figure}
\includegraphics[width=0.45\textwidth]{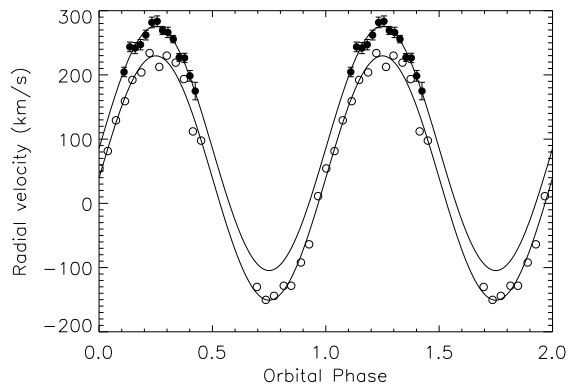} \caption{Measured
radial velocities for the M-dwarf from Bruch (1999) (open circles)
and from this paper (filled circles). Least-squares sinusoidal fits with
independent means but the same semi-amplitude are also shown.
Standard errors for radial velocities from this paper are shown as error bars.
The standard error for the Bruch \& Diaz measurements is assumed to be
  12\,\kms. \label{RedRVFitFig}}
\end{figure}

 The signal-to-noise in the individual spectra is rather low so we created 20
phase-binned spectra over the phase range 0.0 to 1.0, i.e., the weighted
average  of the observed spectra in groups taken at the same orbital phase to
within 0.05 phase units. Our blue spectra show three broad absorption lines
from the white dwarf due to H$_\beta$, H$_\gamma$ and H$_\delta$. Narrow
emission lines from the M-dwarf are seen near the centre of these broad
absorption lines. A few weak absorption features from the M-dwarf,
particularly molecular bands, are present at the red end of the spectrum.

 To measure the radial velocity of the white dwarf from the Balmer lines we
used a two-stage least-squares fit of multiple Gaussian profiles to the
spectra  independently for each Balmer line. The spectra were normalized using
a linear least squares fit to a region of the spectrum either side of the
Balmer line. To achieve a good fit to the absorption line we used the sum of
two Gaussian profiles with the same velocity but independent widths and
depths. A single Gaussian profile was deemed to be a poor fit to these broad
absorption lines.  We used a single Gaussian profile to model the Balmer
emission line and included a first order polynomial in the fit to model the
continuum. The first least-squares fit was used to establish the optimum
widths and depths of the Gaussian profiles by fitting all the spectra
simultaneously. The position of the absorption line was calculated  from the
radial velocity predicted by the expression $V_r = \gamma_{\rm WD} - K_{\rm
WD} \sin( 2\pi\phi)$, where $\phi$ is the phase. A similar expression using
$\gamma_{\rm em}$ and $K_{\rm em}$ was used for the emission line. The values
of $\gamma_{\rm WD}$, $K_{\rm WD}$, $\gamma_{\rm em}$ and $K_{\rm em}$ were
included as free parameters in the fit.  Only data within 1000\kms of the rest
wavelength of the line where included in the fit. The second least-squares fit
we used to measure the radial velocity of the white dwarf in each spectrum
independently. The widths and depths of the Gaussians used to model the white
dwarf absorption were fixed at the values measured in the first step. The free
parameters in the fit were the radial velocity of the white dwarf, the width
and height of the emission line, the radial velocity of the emission line and
the coefficients of the polynomial used to model the continuum. The
width and height of the emission lines were optimised individually for each
spectrum so that the radial velocity measurements of the white dwarf were not
affected by poor fits to the emission lines. The fits to the spectra around
the H$_\beta$ line are shown as a trailed grey-scale plot in
Fig.~\ref{TrailHbetaFig}. Examples of the second least-squares fit to the
spectra at two key phases are shown in Fig.~\ref{SpecFitFig}.  We attempted to
subtract an estimate of the M-dwarf spectrum from the observed spectra to
improve the least-squares fit but found that this had a negligible effect on
the results.

 We used an unweighted least-squares fit to the measured radial velocites for
the white dwarf to determine the optimum values of  $\gamma_{\rm WD}$ and
$K_{\rm WD}$ using the same expression for $V_r$ as above. For the observed
value of the radial velocity at each phase we used  the weighted mean of the
values from the three Balmer lines, excluding one unreliable measurement for
the H$_{\delta}$ line. The resulting fits are given in Table~\ref{RVFitTable}.
Also shown in Table~\ref{RVFitTable} are the values of $\gamma_{\rm em}$ and
$K_{\rm em}$ determined in the same way. The measured radial velocities and
sinusoidal fits are shown in Fig.~\ref{BlueRVFitFig}. It should be noted that
it is difficult to establish an accurate continuum level for the spectra given
the noise in the spectra, the contamination of white dwarf spectrum by the
M-dwarf and the lack of any spectrophotometric standard star observations to
calibrate the instrument response. This is likely to lead to a systematic
error in the radial velocities measured from the broad white dwarf absorption
lines. However, all the spectra will be affected in a similar way so we expect
that this will only affect our estimate of $\gamma_{\rm WD}$. These systematic
effects should have a negligible affect on our estimate of $K_{\rm WD}$. 

\subsection{The spectroscopic orbit of the M-dwarf}

 We measured the radial velocity of the M-dwarf in RR~Cae from our spectra
over the region 8440\,--\,8930\AA\ by cross-correlation. We chose these
wavelength limits to avoid the parts of the spectrum  with strong telluric
absorption. We used a variety of spectra of M-dwarfs taken from the study of
Cenarro et al. (2001) as templates for the cross correlation. We found that
the choice of template has a negligible effect on the results. The results
presented here used the spectrum of the M5-6 dwarf BD~+19$^{\circ}$~5116\,B as
a template spectrum.  These data have a well defined zero-point but insufficient
phase coverage to reliable determine the value of $K_{\rm M}$. Bruch
(1999) provides similar radial velocity measurements for the M-dwarf which
 do have good phase coverage but the zero-point for those measurements is not well
defined. Therefore, we have used a weighted least-squares fit to our radial
velocity measurements and the  radial velocities tabulated by  Bruch (1999)  of
two sinusoidal functions with independent means but a common semi-amplitude
$K_{\rm M}$. The radial velocities were weighted according to their standard
errors. The standard errors for our radial velocity measurements were
calculated from the parabolic fit to the three highest points in the
cross-correlation function plus an additional uncertainty of 2.6\kms\ added in
quadrature to achieve a reduced $\chi^2$ value of 1 in the least-squares fit.
We used the root-mean-square residual (rms) value from a least-squares fit of
a sine wave to the  Bruch radial velocities only to estimate a standard error
of 12\kms\ for those measurements. The results of the least-squares fit to
both data sets are given in Table~\ref{RVFitTable} and are shown in
Fig.~\ref{RedRVFitFig}. 

\subsection{The lightcurve}
\begin{figure}
\includegraphics[width=0.45\textwidth]{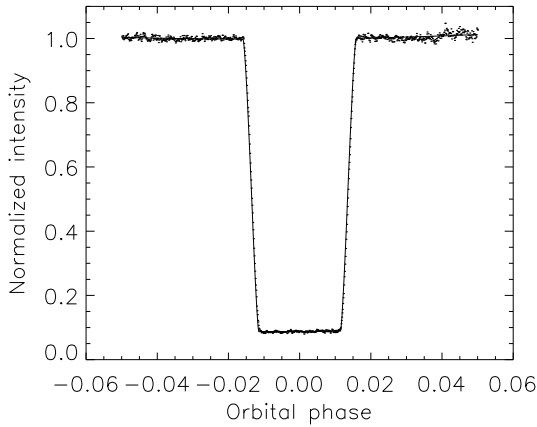}
\caption{\label{LCFitFig}The mean white-light lightcurve of RR~Cae (points)
and a least-squares fit for a uniform disk eclipse (solid line, barely
visible). The mean lightcurve is the average of all our
white-light photometry in the phase range  $-0.05$ to $0.05$ 
re-sampled into 1001 bins.}
\end{figure}

\begin{figure}
\includegraphics[width=0.45\textwidth]{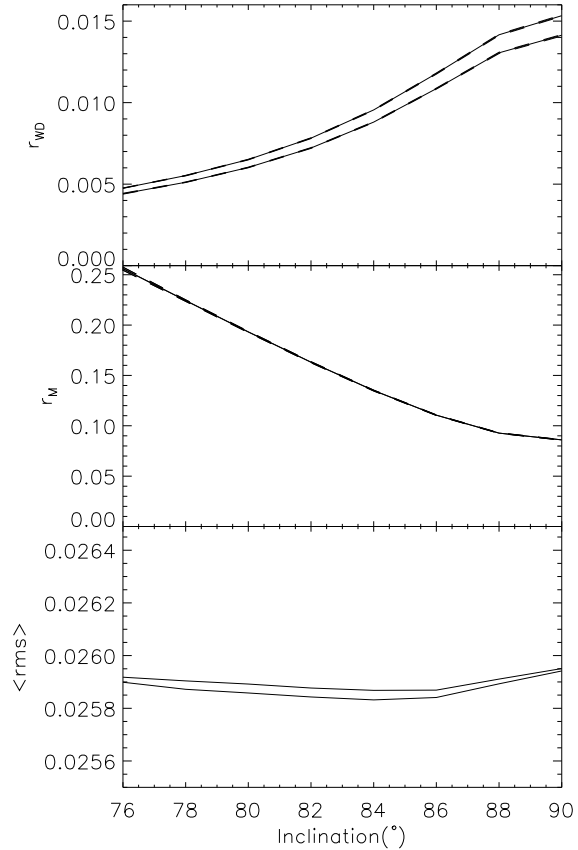}
\caption{Upper panel: The optimum value of $r_{\rm WD}$ as a function of
inclination from a least-squares fit to our white-light photometry using a
linear limb darkening coefficient of $u_1 = 0.35$ (lower solid line) or 0.95 (upper
solid line). The standard error in each curve is indicated using a dashed line
either side of optimum value (barely visible). \newline
Middle panel: The optimum value of $r_{\rm M}$ as a function of
inclination from a least-squares fit to our white-light photometry (solid
line). The standard error in the curve  is indicated using a dashed line
either side of optimum value (barely visible). The effect of the limb darkening
coefficient of the white dwarf is negligible. \newline
Lower panel: The mean rms for the least-squares fits the normalized data for 
individual eclipses. Results are shown for an assumed linear limb darkening
coefficient of $u_1 = 0.35$ (lower curve) and $u_1 = 0.95$ (upper curve).
\label{LCParFig}}
\end{figure}

\begin{figure}
\includegraphics[width=0.45\textwidth]{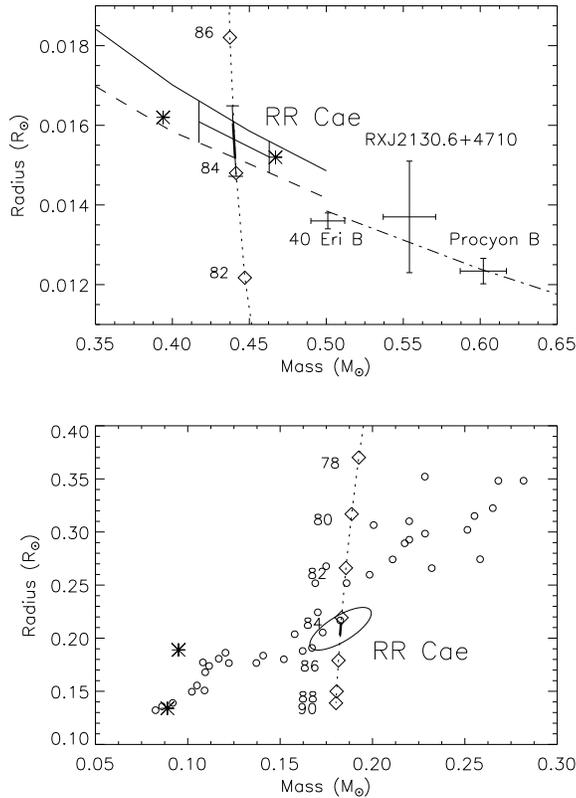}
\caption{Upper panel: The mass-radius relation for RR Cae and other white
dwarfs.   The predicted mass-radius relation from the cooling models of
Benvenuto \& Althaus (1999) is shown as follows: dashed
line -- helium white dwarf with a thin hydrogen layer (X=$10^{-8}$), T$_{\rm
eff}$=7540\,K; solid line -- helium white dwarf with a thick hydrogen layer
(X=$4\times10^{-4}$), T$_{\rm eff}$=7540\,K; dash-dot line -- CO white dwarf
with thin hydrogen layer  (X=$10^{-8}$), T$_{\rm eff}$=10,000\,K. The dotted
line show the relation between mass and radius for RR~Cae inferred from our
lightcurves and the spectroscopic orbit. Diamonds on this line are labelled by
the corresponding inclination in degrees. The thick, solid line shows the
intersection of this relation with the models. The thin capped solid line
extending from this thick line show the  68.3\,per cent confidence interval
for the radius.  The thin, almost-horizontal, solid lines forming the error
bar show the 68.3\,per cent confidence interval for the mass at the best
estimate of the white dwarf mass and radius. The mass and radius of the white
dwarf \rxj\ from Maxted et~al., (2004) is shown. Also shown are the masses and
radii of Procyon~B and 40~Eri~B from Provencal (2002). The estimates for the
mass and radius from Bruch (1999) and Bruch \& Diaz (1998) are shown as
asterisks.
\newline
Lower panel: The M-dwarf component of RR~Cae in the mass-radius plane compared
to other M-dwarfs taken from Clemens et~al. (1998). The thick, solid line and
the dotted line are the mass and radius corresponding to the white dwarf mass
and radius in the panel above. The ellipse is the 68.3\,per cent confidence
region for the M-dwarf mass and radius for the best estimate of the masses 
radius. The estimates for the mass and radius from Bruch (1999)
and Bruch \& Diaz (1998) are shown as asterisks.
\newline
\label{MvRFig}} 
\end{figure}

\begin{table}
\caption{\label{MRTTable} The masses ($M$), radii($R$), effective temperatures
($\Teff$) and surface gravities ($g$) of the stars in RR~Cae. Where a range of
values are given this reflects the range of values calculated from different
cooling models for the white dwarf. The uncertainty quoted is a standard
deviation for a fixed mass and radius  derived from a Monte Carlo analysis.
N.B. the errors in the masses and radii are correlated (see Fig.~\ref{MvRFig}).}
\begin{center}
\begin{tabular}{@{}lrr}
\noalign{\smallskip}
Parameter & \multicolumn{1}{l}{White dwarf} & \multicolumn{1}{l}{M-dwarf} \\
\hline
\noalign{\smallskip}
$M$(\Msolar)&0.440\,$\pm$\,0.023 & (0.182\,--\,0.183)\,$\pm$\,0.012 \\
$R$(\Rsolar)&(0.015\,--\,0.016)\,$\pm$\,0.0004& (0.203\,--\,0.215)\,$\pm$\,0.015 \\
$\Teff$ (K)&$7540\pm 175$&3100\,$\pm$\,100\makebox[0pt][l]{$^{a}$}\\
$\log g$(cgs)   &  (7.67\,--\,7.72)\,$\pm$\,0.06 & (5.04\,--\,5.09)\,$\pm$\,0.04 \\
\noalign{\smallskip}
\hline
\end{tabular}
\end{center}

$^a$ Based on  our estimate of the spectral type  and the
calibration of Leggett (1992).
\end{table}

 The lightcurve of RR~Cae  shows that the white
dwarf is totally eclipsed for 10\,m\,04s and the ingress and egress phases
last 1\,m\,50s (Fig.~\ref{LCFitFig}). The depth of the eclipse varies from 2.6
magnitudes in white-light to 0.25 magnitudes in the I-band.

 For a given inclination, $i$, a single eclipse will give an accurate
measurement of the radii of the stars relative to their separation, i.e.,
$r_{\rm WD} = R_{\rm WD} /a$ and $r_{\rm M} = R_{\rm M} /a$, where $a$ is the
separation of the stars and  $R_{\rm WD}$, $R_{\rm M}$ are the radii of the
white dwarf and the M-dwarf, respectively. Additionally, the depth of the
eclipse gives an accurate measurement of the luminosity ratio, $l_{\lambda}$
at the effective wavelength of the lightcurve, $\lambda$. Good fits to
the eclipse can be found for a wide range of $i$ values. 

 We used least-squares fits to our white-light photometry of the lightcurve
model {\sc ebop} (Etzel 1981; Popper \& Etzel 1981) to measure the functions
$r_{\rm WD}(i)$ and  $r_{\rm M}(i)$ over a grid of inclination values from
$76^{\circ}$ to $90^{\circ}$ in $2^{\circ}$ steps. We fitted each eclipse we
had observed separately and used the standard error of the mean for the
results to estimate the standard error in $r_{\rm WD}(i)$ and  $r_{\rm M}(i)$
at each value of $i$.  The data from JD\,2450700 were excluded from the
analysis because they give discrepant results, for reasons that will be
discussed below. The value of $r_{\rm WD}$ depends slightly on the limb
darkening values adopted for the the white dwarf. Atmospheres of DA white
dwarfs usually consist of pure hydrogen. However, 
in the case of RR Cae the presence of a number of metal lines indicates 
significant contamination by heavy elements, which alters the structure of 
the atmosphere and thus the limb darkening coefficient. A detailed 
determination of the abundance mix is beyond the scope of this 
paper. For this reason we consider a wide range 
of linear limb darkening coefficients from $u_1=0.35$ to $0.95$ in
steps of 0.2. The values of $r_{\rm WD}(i)$ and  $r_{\rm M}(i)$ we derive are
shown in Fig.~\ref{LCParFig}. It is clear that the standard errors contribute
negligibly to the uncertainty in  $r_{\rm WD}(i)$ and  $r_{\rm M}(i)$ and that
the  value of $r_{\rm M}(i)$ does not depend on $u_1$. We also show the mean
of the rms values for the least-squares fits to the lightcurves to demonstrate
that there is very little difference in the goodness-of-fit for a wide range
of inclination.

\subsection{The effective temperature of the white dwarf\label{sb2}}
\begin{figure}
\includegraphics[width=0.45\textwidth]{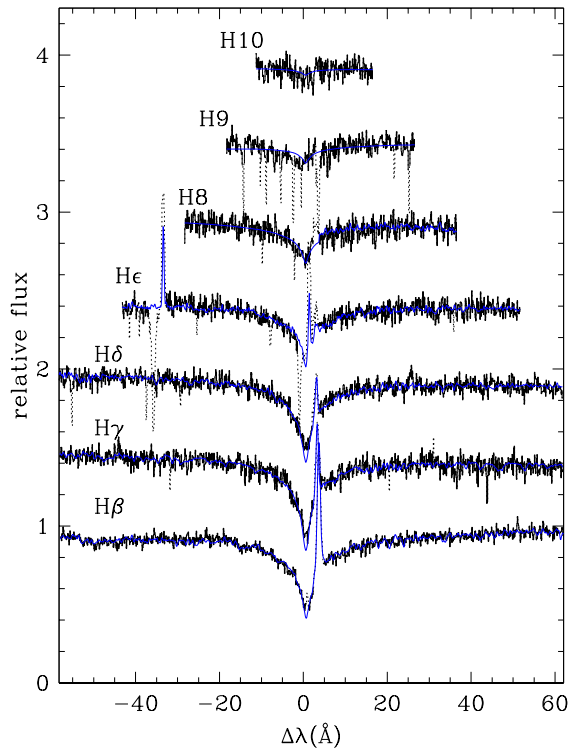}
\caption{Determination of the white dwarf parameters by a model
   atmosphere fit of the composite spectrum of RR~Cae (spectrum taken
   at HJD 2\,451\,889.7315 shown here). The M-dwarf spectrum is
   computed for parameters determined in a previous step (see
   text). Dotted lines indicate M star emission lines and photospheric
   white dwarf metal absorption lines excluded from the fit.
   \label{SPYSpecFitFig}}
\end{figure}

Bragaglia et al.\ (1995) estimated the white dwarf parameters from
model atmosphere fits to the Balmer lines. However, that analysis is
seriously hampered by the uncorrected contamination of the spectrum by
the light of the M-dwarf companion. Here we present an analysis which
overcomes this shortcoming.

 RR~Cae was observed as part of the Supernovae Type Ia Progenitor Survey (SPY
-- Napiwotzki et al. 2001). Spectra were taken with the high-resolution
echelle spectrograph UVES attached to the UT2 telescope of VLT. The SPY setup
used for the UVES observations provides almost complete spectral coverage from
3200\,\AA\ to 6650\,\AA\ with a resolution of about 0.3\,\AA. Further details
of the observations and data reduction are given in Napiwotzki et al.\ (2001).

 The analysis was performed with the {\sc fitsb2} code, which is designed to
perform a model atmosphere analysis of double lined binary systems in which
both components are visible (Napiwotzki et al.\ 2004). Radial velocities and
stellar parameters of both components can be treated as free parameters. The
spectrum of the white dwarf in RR~Cae was modelled with a grid of spectra
computed with the model atmosphere code of Detlev Koester described in Finley
et~al. (1997). The spectrum of the M-dwarf was modelled using template spectra
from the library of stellar spectra of Cincunegui \& Mauas (2004). This
library contains flux calibrated high resolution spectra. Spectral type is
used as the fit parameter for the M-dwarf.

 Relative flux contributions at 6700\,\AA\ from the white dwarf and the
M-dwarf were fixed at the value $F_{\mathrm{WD}}/F_{\mathrm{M
star}}=1.68$ derived by Bruch \& Diaz (1998) from their light curve
analysis.  As a first step we made use of the full spectral range from
the Balmer lines to H$\alpha$ to constrain the spectral type and radial
velocities of the M-dwarf. White dwarf parameters were allowed to vary
as well during this phase. We estimate the spectral type of the M-dwarf to be
about M4. For the final fit to spectrum we restricted the fit to the Balmer
lines H$\beta$ and higher with M-dwarf parameters fixed at the previously
measured values. We used two spectra taken at epochs HJD=2\,451\,889.7315 and
2\,451\,891.6927 for this analysis. The final fit is compared to one of the
observed spectra in Fig.~\ref{SPYSpecFitFig}.

The resulting white dwarf parameters are $T_{\mathrm{eff}}=7540$\,K$\pm
175$\,K and $\log g = 7.97$.  The standard error in  $T_{\mathrm{eff}}$ is
taken to be 2.3\,per\,cent (Napiwotzki, Green \& Saffer 1999).
The uncertainty in the value of $\log g$ is discussed in
Section~\ref{Discussion}. The new temperature is higher than the Bragaglia
et~al. (1995) value, indicating a somewhat smaller cooling age than assumed
previously. 

\subsection{Masses and radii of the stars}
For each value of $i$ and $u_1$ we can combine the value of $r_{\rm WD}$ and
$r_{\rm M}$ with  the spectroscopic orbits in Table~\ref{RVFitTable} to derive
the masses and radii for the stars. To estimate a nominal value of  $u_1$ we
refer to the linear limb darkening coefficients calculated by Claret (2000)
for the Johnson U,B and V filters for $\Teff =  6750,7000,7250$K,  $\log g =
4.5,5.0$ and $\log[M/H]=-4.5$ (where $[M/H]$ is the heavy metal abundance
relative to the Sun). Extrapolating to the $\log g$ value of RR~Cae, we
estimate that the value of $u_1$ likely to be about $0.65\pm0.15$. We have
adopted a large standard error on this estimate to reflect the uncertainty due
to extrapolating over a large range of $\logg$ and the poorly defined
band-pass for our white-light photometry. Fortunately, the actual value of
$u_1$ adopted has a negligible effect on our results.

 The masses and radii we derive for the stars as a function of inclination are
shown in Fig.~\ref{MvRFig}. Also shown in Fig.~\ref{MvRFig} are the
mass-radius relations for white dwarfs with various compositions from the
cooling models of Benvenuto \& Althaus (1999). As we have no further useful
observational constraint on the inclination we use the range of radii for the
white dwarf predicted by these models to find the likely range of radii for
the M-dwarf. The range of values we derive for the radii of the stars are
given in Table~\ref{MRTTable} and are shown as thick, solid lines in
Fig.~\ref{MvRFig}. We see that the masses of the stars are well determined and
the main source of uncertainty in the white dwarf radius is the unknown mass
of the surface hydrogen layer. We use the value of $K_{\rm M}$ measured from
the absorption line spectrum of the M-dwarf to estimate the masses because the
emission lines may arise from regions that are not uniformly distributed over
the surface of the M-dwarf. Although  the value of $K_{\rm em}$ is more
precise than  $K_{\rm M}$ it may not accurately reflect the orbital motion of
the centre-of-mass of the M-dwarf.
 
 We have used a Monte Carlo technique to estimate the uncertainties on the
masses and radii we have measured. For each of 1000 trials we select random
deviates from a normal distribution for every quantity that is used to
calculate the masses and radii, i.e., $K_{\rm WD}$, $K_{\rm M}$, $u_1$,
$r_{\rm WD}(i)$ and  $r_{\rm M}(i)$. We use two values of $i$ corresponding to
the extreme values of $R_{\rm WD}$ found for cooling models with thin and
thick hydrogen layers. We then calculate and store the masses and radii of the
stars for every trial so that we can calculate their standard deviations and
the correlation between the parameters. The standard errors derived are given
in Table~\ref{MRTTable}. The joint confidence region for the mass and radius
of the M-dwarf are shown as an ellipse in the lower panel of
Fig.~\ref{MvRFig}.

\section{Discussion}\label{Discussion}

\begin{figure*}
\includegraphics[width=0.95\textwidth]{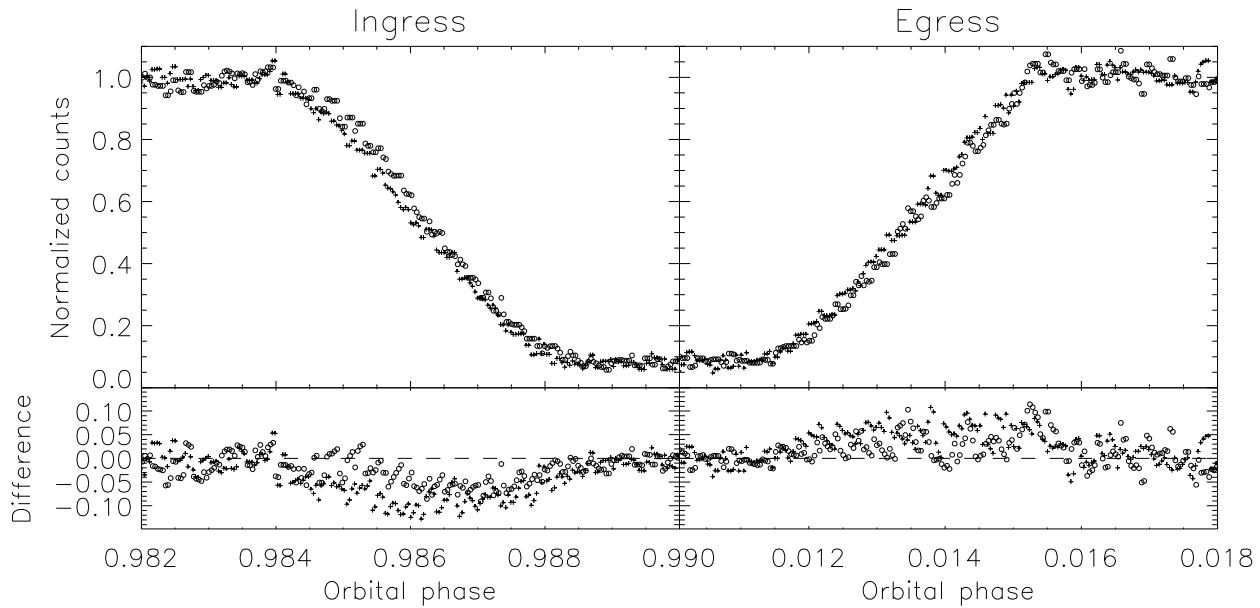}
\caption{\label{EclipseFig}A comparison of two eclipses observed on
consecutive nights with the SAAO 1-m telescope using white-light photoelectric
photometry. The data are from the nights 03 Jan 1995 (crosses) and 04
Jan 1995 (open circles). The lightcurves have been normalized independently
to a mean out-of-eclipse level of 1 and smoothed using a three point median
filter. The difference between these lightcurves and the average lightcurve
shown in Fig.~\ref{LCFitFig} is also shown below each portion of the
lightcurve.}
\end{figure*}

 The mass we have derived for the M-dwarf in RR~Cae is very different from the
much lower masses given by Bruch (1999) or Bruch \& Diaz (1998). Our analysis
is more reliable than those studies because we have measured a dynamical
mass ratio for the stars from the spectroscopic orbits. In contrast to those
studies, we find that the radius of the M-dwarf is consistent with
its mass (Fig.~\ref{MvRFig}) and our estimate of the spectral type (Leggett
1992). We fully agree with the statement of  Bruch (1999) that the uncertainty
ranges quoted in that study ``do not take into account the correlation between
parameters. Thus, they must not be mistaken as errors of the corresponding
parameters!''.

 In principle, the gravitational redshift of the white dwarf can be used as an
additional constraint to determine a unique solution for the masses and radii
of the stars independently of the white dwarf cooling models. The
gravitational redshift of the white dwarf is expected to be about 17\,\kms. This
is clearly not consistent with the values of $\gamma_{\rm WD}$ and
$\gamma_{\rm M}$ we have measured. We believe this is due to a systematic
error in  our value of $\gamma_{\rm WD}$ (see Section~\ref{SBWD}). The UVES
spectra discussed in Section~\ref{sb2} yield two radial velocities per star
which we estimate have uncertainties of a few \kms. Since the phase of the
observations is known this is sufficient to estimate the values of
$\gamma_{\rm WD}$, $\gamma_{\rm M}$, $K_{\rm WD}$ and $K_{\rm M}$. We find that
the values of $K_{\rm M}$ and $K_{\rm WD}$ are consistent with the values
shown in Table~\ref{RVFitTable} to within a few \kms. The difference
$\gamma_{\rm WD}-\gamma_{\rm M}\approx 10$\kms\ is closer to the expected value
than the value in Table~\ref{RVFitTable}.

 The surface gravity of the white dwarf, $\logg$, derived from the
analysis of the spectrum is larger than the range computed from the orbital
solution and lightcurve analysis given in Table~\ref{MRTTable} by about
0.3\,dex. Systematic errors of this type in the value of $\logg$ for cool
white dwarfs have been noted by several authors (Kleinman et~al. 2004;
Bergeron, Wesemael, \& Fontaine, 1992; Bergeron et~al. 1990) but the reasons
for this effect are not clear. It has been suggested that the effect is a
result of an increase in the mean mass of the samples  towards cooler
temperatures. This clearly cannot apply in the case of RR~Cae. The mass
implied by the values of $T_{\mathrm{eff}}$ and $\log g$ derived in
Section~\ref{sb2} is about 0.57\Msolar\ for a cooling model with a hydrogen
layer thickness of $10^{-4}$ (Benvenuto \& Althaus 1999).

 We have calculated the timescale for circularisation by tidal forces 
of the orbit in RR~Cae and find that it is only a few thousand years. For this
reason, we discount the suggestion by Bruch (1999) that there is a significant
eccentricity in the orbit. We do not speculate here  on the origin of the
feature in the R-band lightcurve offset from phase 0.5 which he suggests may
be a secondary eclipse, but note that our radius measurements for the stars in
RR~Cae predict that the secondary eclipse has a depth of 7\,milli-magnitudes or
less.

 We have re-calculated the past and future evolution of RR~Cae using the
analysis described by Schreiber \& G\"{a}nsicke (2003) but with our new, more
accurate values for the masses and radii of the stars.  The  cooling age,
$t_{\rm cool}$, of RR~Cae derived by interpolating the cooling tracks of Wood
(1995) is $\log(t_{\rm cool}/y) =8.94$. This is the time since RR~Cae emerged
from the common-envelope phase, at which time the orbital period is estimated
to have been 0.31\,--\,0.32\,d depending on the prescription used to model the
angular momentum loss due to a magnetic stellar wind in this interval.
 The
continued loss of angular momentum will shrink the Roche lobe of the M-dwarf
to the point where mass transfer will start from the M-dwarf to the white
dwarf through the inner Lagrangian point. RR~Cae will then be a cataclysmic
variable star (CV). This will occur at an orbital period $P_{\rm
sd}=0.080$\,--\,$0.086$\,days. Our estimate of the time before Roche lobe
overflow occurs, $t_{\rm sd}$, also depends on the prescription for angular
momentum loss used and varies from $\log(t_{\rm sd}/y) = 9.94$ to $\log(t_{\rm
sd}/y) = 10.3$. These values are not much changed from those presented by
Schreiber \& G\"{a}nsicke, but the changes are sufficient to alter two of
their conclusions. Firstly, the revised value of $P_{\rm sd}$ is at the lower
limit of the range 2\,--\,3\,h in which very few non-magnetic CVs are observed
(the CV period gap) rather than within this period range. Secondly, the value
of $t_{\rm sd}$ for RR~Cae may be consistent with the definition of a pre-CV
used by Schreiber \& G\"{a}nsicke, i.e., post-common envelope binaries that
evolve into a semi-detached configuration in less than the Hubble-time.

 We have also used the prescriptions for angular momentum loss described in
Schreiber \& G\"{a}nsicke to calculate the current rate of period change,
$\dot{P}/P$, for RR~Cae. We find that for classical magnetic braking, (Verbunt
\& Zwaan 1981) $\dot{P}/P \approx 5\times10^{-14}$ and for revised magnetic
braking, (Pinsonneault, Andronov \& Sills 2002) $\dot{P}/P \approx
1.4\times10^{-13}$. In either case, we would not expect to detect any period
change given the accuracy of our eclipse timings. Continued monitoring of the
eclipse times of RR~Cae will, in principle, enable us to distinguish between
these two prescriptions for magnetic braking, although the predicted change in
period is only sufficient to change the predicted time of minimum by about 1
second over 10 years. 

 A concerted observing campaign will be required to reliably measure such a
small change in the orbital period of RR~Cae, particularly in the presence of
the additional source of noise identified from our white-light photometry. We
have considered several possibilities for the source of this additional error.
There is no evidence for systematic errors in the photometry at phases outside
eclipse large enough to account for the changes in eclipse timings were
similar systematic errors to occur during the eclipse. Careful inspection of
the residuals shows that the ingress and egress phases are not affected
equally, so the external noise source is not errors in the timing of the
observations. We looked for the influence of flares from the M-dwarf on the
eclipse timings but did not see any flare-like features in the lightcurves of
any significance.

 Watson \& Dhillon (2004) describe how the Wilson effect can perturb the
eclipse timings of eclipsing binaries like RR~Cae by a few seconds. Their
simulation of the Wilson effect shows that the cause of the change in eclipse
timing is a shift of the whole ingress or egress phase of the eclipse. In
Fig.~\ref{EclipseFig} we compare two eclipses taken on subsequent nights, both
of which are a few seconds early compared to our linear ephemeris but by
clearly different amounts (2.7\,s and 6.1\,s). We see that the reason for this
difference is a change of shape of the ingress phase, not a simple shift. This
is particularly noticable in the difference between these lightcurves and the
mean lightcurve in the phase range 0.985\,--\,0.986. There is also an overall
shift to the time of ingress and egress compared to the mean lightcurve. We
conclude that the Wilson effect may contribute to the variability in the times
of mid-eclipse but some other phenomenon is also contributing to these
variations.

We also considered the possibility that the white dwarf in RR~Cae has star
spots due to a magnetic field similar to those seen in the cool white dwarf
WD\,1953$-$011 (Maxted 2000; Brinkworth et~al. 2005).  The detection
of sharp metal lines in the spectrum of RR~Cae by Zuckerman et~al. (2003)
would appear to rule out this possibility because we would expect these lines
to show strong Zeeman splitting in this scenario. In order to produce the
variability we observe in RR~Cae the viewing angle between the observer and
the hypothetical spot must change. At some orientations the line of sight will
be along the magnetic field lines, in which case the longitudinal field
produces a split pair of circularly polarized $\sigma$ components in the
spectral line and suppresses the central linearly polarized $\pi$ component
(Rutten 1996), i.e, the orientation of the magnetic field is favourable for
the detection of Zeeman splitting. Strong Zeeman splitting of the lines will
not be apparent if the orbital and spin axes of the white dwarf are aligned
with the magnetic field, but this orientation does not give rise to any
rotational modulation of the brightness. 

 Rather than a dark spot due to a strong magnetic field, it may be that bright
regions on the white dwarf due to accretion may contribute to the variability
of the eclipse timings.  Accretion from the solar-like wind of the M-dwarf
is thought to  be responsible for the presence of sharp metal lines in the
spectrum of RR~Cae (Zuckerman et~al. 2003). This hypothesis can be tested by
looking for correlations between the overall apparent brightness of the white
dwarf and the presence of distortions to the eclipse lightcurve. If the
distortions to the lightcurve seen in Fig.~\ref{EclipseFig} are interpreted as
changes in brightness of the white dwarf then the overall brightness changes
at blue wavelengths are expected to be quite large (about 5 per cent). The
mass transfer rate onto the white dwarf required to generate an accretion
luminosity equal to  5 per cent of the white dwarf luminosity is
$3\times10^{-14}$\Msolar/y. Debes (2006) has calculated the accretion
rate onto the white dwarf implied by the observed abundance of Ca\,II in the
atmosphere of RR~Cae and the corresponding mass loss rate from the M-dwarf
assuming Bondi-Hoyle accretion. They find a mass accretion rate $\dot
M=4\times10^{-16}$\,\Msolar/y and a mass loss rate $\dot {M_{\rm RD}}
=6\times10^{-16}$\,\Msolar/y. This can be compared to the upper limit to the
mass loss rate of $\le 2\times10^{-15}$\Msolar/y for the M5.5 dwarf Proxima
Cen (Wood et al. 2001). If accretion onto the white dwarf is responsible for
the distortions to the lightcurve then the M-dwarf in RR~Cae is much more
active than Proxima Cen and/or the mass transfer onto the white dwarf is very
non-uniform.  Intriguingly, a footnote to a table in Zuckerman et~al.  for
RR~Cae notes ``accretion disc/hot spot on WD?'', though it is not clear what
leads the authors to make this suggestion.

The DAB white dwarf HS\,0209+0832 appears to be an example of a WD with no 
significant magnetic field that shows variations of chemical abundances over
the surface (Heber et al. 1997). RR Cae is much cooler than HS\,0209+0832 so
the hydrogen is almost completely neutral.  In these conditions trace metals
are an important sources of free electrons and opacity. Variations in
chemistry across the surface of RR Cae similar to those seen in HS\,0209+0832
may result in local changes in the atmospheric structure that may be related
to the variability in eclipse shape that we have seen. In summary, none of the
explanations above is entirely satisfactory but some progress may be made by
looking for variability in the strength of the metal lines in RR~Cae.

 Bruch (1999) found that the equivalent width of the H$\alpha$ emission lines
is variable by about 50\,per\,cent with a tendency for the line be brighter at
phases close to eclipse. This is exactly the opposite of the behaviour
expected if the emission line is due to irradiation of the M-dwarf by the
white dwarf. Our measurements of the equivalent width of the Balmer emission
lines are not as accurate as those of Bruch because the lines are weaker in
the higher Balmer lines. Some variability is apparent, but this is not
coherent with orbital phase, e.g., a flare-like brightening of the H$\beta$
line is apparent around phase 0.3 in Fig.~2. This enforces the conclusion of
Bruch that the emission lines are the result of chromospheric activity on the
M-dwarf.

\section*{Acknowledgments}
Based on observations obtained at the South African Astronomical Observatory.
 PFLM and LMR were supported by a PPARC post-doctoral grants. The calculation
of the eclipsed area for our simple lightcurve model is taken from a {\sc
basic} code written by Dan Bruton of the Department of Physics \& Astronomy,
Stephen F. Austin State University, Texas.  We thank Detlev Koester for
providing us with a grid of model atmospheres. We thank the referee for
comments on the manuscript that have helped us to clarify several points in
the text.

\label{lastpage}
\end{document}